\documentclass[aps,prd,superscriptaddress,noshowpacs,nofootinbib,amssymb,balancelastpage,preprintnumbers]{revtex4}

\usepackage{graphicx}
\usepackage{wrapfig}
\usepackage{amssymb}
\usepackage{amsmath}
\usepackage{color}
\usepackage{hyperref}

\newcommand{\vev}[1]{{\langle #1 \rangle}}

\begin{document}
\renewcommand{\thefootnote}{\#\arabic{footnote}} 
\setcounter{footnote}{0}
%
\preprint{KEK-TH-2253}

\title{Hiding neutrinoless double beta decay in the minimal seesaw mechanism}

\author{Takehiko Asaka}\thanks{{\tt asaka@muse.sc.niigata-u.ac.jp}}
\affiliation{Department of Physics, Niigata University, Niigata 950-2181, Japan}
\author{Hiroyuki Ishida}\thanks{{\tt ishidah@post.kek.jp}}
\affiliation{KEK Theory Center, IPNS, Tsukuba, Ibaraki 305-0801, Japan}
\author{Kazuki Tanaka}\thanks{{\tt tanaka@muse.sc.niigata-u.ac.jp}}
\affiliation{Graduate School of Science and Technology, Niigata University, Niigata 950-2181, Japan}


\begin{abstract}
We present a possibility that the neutrinoless double beta decay 
can be hidden in the minimal seesaw mechanism 
where the standard model is extended by two right-handed neutrinos 
which have a hierarchical mass structure. 
In this framework, the lepton number is violated due to the massive Majorana neutrinos. 
Especially, we investigate the case that 
the heavier right-handed neutrino is sufficiently heavy to decouple from the decay 
while the lighter one is lighter enough than the typical Fermi-momentum scale of nuclei 
and gives a sizable contribution to the decay. 
Under the specific condition on mixing elements, 
the lighter right-handed neutrino can give a significant destructive contribution 
which suppresses or even hides to the effective mass of the neutrinoless double beta decay. 
In this case, the flavor structure of the mixing element of the lighter right-handed neutrino with ordinary neutrinos 
is predicted depending on the Majorana $CP$ violating phase of active neutrinos.
\end{abstract}
\maketitle

\section{Introduction}
Neutrino oscillation experiments have been developed so far
and have provided many properties of three active neutrinos, 
including mass square differences, mixing angles, 
and even $CP$ violating phase which has been getting to be revealed recently. 
Unfortunately, all the fascinated neutrino oscillation experiments 
can give no hint of whether neutrinos are Dirac or Majorana fermions,
which is one of the most interesting missing pieces of neutrinos. 
One promising possibility to attack this issue is finding the phenomenon 
of the neutrinoless double beta ($0 \nu \beta \beta$) decay. 
(See as a theoretical review, for instance~Ref.~\cite{Pas:2015eia}.) 
It violates the lepton number by two units and shows a clear signal 
for physics beyond the Standard Model (SM). 
The decay can be mediated by massive Majorana neutrinos and the rate is
characterized by the so-called effective mass $m_{\rm eff}$ of Majorana neutrinos,
which has been constrained by various $0 \nu \beta \beta$ decay experiments until today~%
~\cite{Arnaboldi:2002te,Umehara:2008ru,Barabash:2010bd,Gando:2012zm,Agostini:2013mzu,Albert:2014awa,Andringa:2015tza,Arnold:2015wpy,Arnold:2016ezh,KamLAND-Zen:2016pfg,Elliott:2016ble,Arnold:2016qyg,Arnold:2016bed,Agostini:2017iyd,Aalseth:2017btx,Albert:2017owj,Alduino:2017ehq,Agostini:2018tnm,Azzolini:2018dyb,Arnold:2018tmo,Adams:2019jhp,Alvis:2019sil,Agostini:2019hzm,Azzolini:2019tta,Alenkov:2019jis,Anton:2019wmi}.

The seesaw mechanism~%
\cite{Minkowski:1977sc,Yanagida:1979as,Yanagida:1980xy,Ramond:1979,GellMann:1980vs,Glashow:1979,Mohapatra:1979ia} by introducing right-handed neutrinos with Majorana masses 
is the one of the most attractive scenarios 
for explaining the origin and the observed smallness of neutrino masses. 
In this case the lepton number is violated by the Majorana masses 
and the $0 \nu \beta \beta$ decay is possible to occur. 
The effective mass is expressed in terms of neutrino masses, 
mixing angles, and $CP$ violating phases.
In the effective SM with three massive neutrinos, 
there is a possibility 
where the effective mass is highly suppressed or even vanishes 
by tuning the lightest active neutrino mass (see, {\it e.g.,}~\cite{Pas:2015eia}).

When right-handed neutrinos are much lighter than the unification scale $\sim 10^{16}~{\rm GeV}$, 
or even lighter than the weak scale $\sim 100~{\rm GeV}$, 
they can give a sizable contribution to $m_{\rm eff}$ in addition to active neutrinos' one. 
In such cases, the masses and mixing elements of right-handed neutrinos 
must be chosen appropriately without conflicting with the experimental limits on $m_{\rm eff}$. 
When all right-handed neutrinos are lighter than a typical scale of Fermi momentum
of a nucleus $\Lambda_\beta$ ($\sim \mathcal{O} (100)$~MeV), 
the contributions of active neutrinos and heavier ones 
exactly cancel out each other due to the intrinsic property of the seesaw mechanism~\cite{Blennow:2010th}. 
Further, it is shown that, when right-handed neutrinos are degenerate, 
$m_{\rm eff}$ becomes smaller than the one solely from active neutrinos~\cite{Asaka:2011pb}. 
Although such right-handed neutrinos are attractive to realize 
the baryogenesis via the oscillation mechanism~\cite{Akhmedov:1998qx,Asaka:2005pn}, 
there is no concrete reason to constrain ourselves to keep the degeneracy in general.

In this paper, we present another possibility to suppress the $0 \nu \beta \beta$ decay 
by hierarchical right-handed neutrinos. 
As the simplest example, we consider the extended SM with two right-handed neutrinos
where one is sufficiently heavier than $\Lambda_\beta$ to decouple from the system 
while the other is lighter than $\Lambda_\beta$ giving a destructive contribution to the decay rate.
It is shown that $m_{\rm eff} = 0$ is possible 
due to the exact cancellation of the contributions between active neutrinos 
and the lighter right-handed neutrino 
if the mixing elements are chosen to be specific values. 
We then discuss the impacts of this cancellation conditions, 
especially, on the Majorana $CP$ violating phase 
and the mass hierarchy of active neutrinos.

\section{Seesaw model with two right-handed neutrinos}
We consider here the simplest extension of the SM 
to explain the observed neutrino masses 
by adding two right-handed neutrinos%
\footnote{
The extension to the case with three right-handed neutrinos is straightforwardly possible. 
However, since the number of parameters is increased, the impacts discussed below would be blurred. 
This issue is beyond our scope.}
$\nu_{R I}$ ($I=1,2$) 
\begin{align}
  {\cal L}_\nu
  = 
  i \overline{\nu_{RI}} \gamma^\mu \partial_\mu \nu_{RI}
  -\left(
  F_{\alpha I} \, \overline{\ell_{\alpha}} \, \Phi \, \nu_{RI}
  +
  \frac{M_I}{2} \, \overline{\nu_{RI}^c} \, \nu_{RI}
  +
  H.c.
  \right)
  \,,\label{Eq:Lag}
\end{align}
where $\Phi$ and $\ell_\alpha$ ($\alpha = e, \mu,\tau$) 
are the Higgs and lepton doublets of the weak SU(2).  
Neutrino Yukawa coupling constants and Majorana masses 
of right-handed neutrinos are denoted by 
$F_{\alpha I}$ and $M_I$, respectively.  
Here and hereafter, we work in the basis 
where the Yukawa coupling matrix of charged leptons 
and the Majorana mass matrix of right-handed neutrinos are diagonal.

The electroweak symmetry breaking gives the neutrino masses
of Dirac type $[M_D]_{\alpha I} = F_{\alpha I} \langle \Phi \rangle$
in addition to the Majorana type $M_I$.  
When $|[M_D]_{\alpha I}|\ll M_I$, the seesaw mechanism 
for neutrino masses is realized. 
In addition to massive active neutrinos $\nu_i$ ($i=1,2,3$)
there are heavy neutrinos $N_I$ 
with masses $M_I$ which almost correspond to right-handed neutrino states 
(we simply call them as right-handed neutrinos from now on).
These states take part in weak gauge interactions 
through the mixing as
\begin{align}
  \nu_{L \alpha} = U_{\alpha i} \, \nu_i + \Theta_{\alpha I} N_I^c \,,
\end{align}
where $U_{\alpha i}$ is the mixing matrix of active neutrinos~\cite{Pontecorvo:1958,Maki:1962mu} 
while the mixing elements of $N_I$ are given by $\Theta_{\alpha I} = [M_D]_{\alpha I} \, M_I^{-1}$.

Based on the parametrization proposed by Casas and Ibarra~\cite{Casas:2001sr,Abada:2006ea}, 
the Yukawa couplings are written as
\begin{align}
  \label{eq:F}
    F = \frac{i}{\vev{\Phi}} \,
    U \, D_\nu^{1/2} \, \Omega \, D_N^{1/2} \,.
\end{align}
Here $D_\nu = \mbox{diag}(m_1, m_2, m_3)$ is the diagonal mass matrix 
of active neutrinos.  In the considering case, 
the lightest active neutrino is massless, and then
$m_3 > m_2 > m_1 = 0$ for the normal hierarchy (NH) case 
and
$m_2 > m_1 > m_3 = 0$ for the inverted hierarchy (IH) case.
$D_N = \mbox{diag}(M_1,M_2)$ is the mass matrix of right-handed neutrinos. 
The mixing matrix of active neutrinos 
is expressed as
\begin{widetext}
\begin{align}
  U = 
  \left( 
    \begin{array}{c c c}
      c_{12} c_{13} &
      s_{12} c_{13} &
      s_{13} e^{- i \delta} 
      \\
      - c_{23} s_{12} - s_{23} c_{12} s_{13} e^{i \delta} &
      c_{23} c_{12} - s_{23} s_{12} s_{13} e^{i \delta} &
      s_{23} c_{13} 
      \\
      s_{23} s_{12} - c_{23} c_{12} s_{13} e^{i \delta} &
      - s_{23} c_{12} - c_{23} s_{12} s_{13} e^{i \delta} &
      c_{23} c_{13}
    \end{array}
  \right)
  \times
  \mbox{diag} 
  ( 1 \,,~ e^{i \eta} \,,~ 1) \,,
\end{align}
\end{widetext}
with $s_{ij} = \sin \theta_{ij}$ and $c_{ij} = \cos \theta_{ij}$.  
$\delta$ and $\eta$ are the Dirac and Majorana $CP$ violating phases, respectively.
The $3 \times 2$ matrix $\Omega$ can be expressed as
\begin{align}
  \Omega =
	\left\{
		\begin{array}{l l}
				\left(
    \begin{array}{c c}
      0 & 0 \\
      c_\omega & - s_\omega \\
      \xi s_\omega & \xi c_\omega
    \end{array}
  \right) &\hspace{2mm} \mbox{for the NH case}
			\\[6ex]
  \left(
    \begin{array}{c c}
      c_\omega & - s_\omega \\
      \xi s_\omega & \xi c_\omega \\
      0 & 0 
    \end{array}
  \right) &\hspace{2mm} \mbox{for the IH case}			
		\end{array}
	\right. \,,
\end{align}
where $s_\omega =\sin \omega$ and $c_\omega = \cos \omega$, respectively. 
$\xi = \pm 1$ is  sign parameter 
and $\omega$ is a complex parameter, {\it i.e.},
$\omega = \omega_r + i \omega_i$.
Further, we introduce
\begin{align}
    X_\omega = \exp [ \omega_i ] \,,
\end{align}
since it represents the overall strength of the Yukawa couplings
(see, e.g., the discussion in Ref.~\cite{Asaka:2011pb}).
In practice, the Yukawa couplings scale as
 $F \propto X_\omega$
or $X_\omega^{-1}$ for $X_\omega \gg 1$ or $\ll 1$.

Throughout this analysis, we choose the convention in which $\xi$ is selected to be positive,
and fix $\theta_{ij}$ and $\delta$ in the mixing matrix $U$ 
to be the central values of the latest global fit
of neutrino oscillation data~\cite{Esteban:2018azc,nufit}. 

\section{Neutrinoless double beta decay}
In the considering model, the effective mass in the 
0$\nu \beta \beta$ decay is given by 
\begin{align}
	m_{\rm eff} =
	m_{\rm eff}^\nu + m_{\rm eff}^{N} \,,
\end{align}
where the contribution from active neutrinos is
\begin{align}
	m_{\rm eff} = \sum_i \, U_{ei}^2 \, m_i \,.
\end{align}
Note that, since only two right-handed neutrinos are introduced, 
it is impossible to cancel the effective mass from active neutrinos by tuning the lightest neutrino mass.
The contribution from right-handed neutrinos is 
\begin{align}
	m_{\rm eff}^{N} = \sum_I  f_\beta (M_I) \, \Theta_{e I}^2 \, M_I \,.
\end{align}
The function $f_\beta$ represents the suppression by the propagator effect
of right-handed neutrinos, and we use the approximate formula
\begin{align}
	f_\beta (M_I) = \frac{\Lambda_\beta^2}{\Lambda_\beta^2 + M_I^2} \,,
\end{align}
where $\Lambda_\beta$ is a typical scale of Fermi momentum of a nucleus 
which is evaluated as a few hundred MeV varied depending on nucleus and modelings~\cite{Faessler:2014kka,Hyvarinen:2015bda,Menendez:2017fdf,Barea:2015kwa}.

We consider the case where $M_1 < \Lambda_\beta \ll M_2$ 
and take $f_\beta(M_1) = 1$ and $f_\beta (M_2) =0$ approximately. 
Thus, the effective mass becomes independent of $\Lambda_\beta$. 
In this case the effective neutrino mass is expressed as
\begin{align}
	m_{\rm eff}
	=
	\left\{
		\begin{array}{l l}
			\left( s_\omega \, U_{e2} m_2^{1/2} - c_\omega U_{e3} m_3^{1/2}
			\right)^2 &\hspace{2mm} \mbox{for the NH case}
			\\[4ex]
			\left( s_\omega \, U_{e1} m_1^{1/2} - c_\omega U_{e2} m_2^{1/2}
			\right)^2 &\hspace{2mm} \mbox{for the IH case}			
		\end{array}
	\right. \,.
\end{align}
Importantly, we find out that the effective mass vanishes 
if the complex parameter $\omega$ satisfies 
\begin{align}
	\tan \omega 
	=
	\left\{
		\begin{array}{l l}
			\displaystyle
			\frac{U_{e3} m_3^{1/2}}{U_{e2} m_2^{1/2} }
			&
			~~~\mbox{for the NH case}
			\\[4ex]
			\displaystyle
			\frac{U_{e2} m_2^{1/2}}{U_{e1} m_1^{1/2} }
			&
			~~~\mbox{for the IH case} 
		\end{array}
	\right. \,.\label{Eq:vanishing-cond}
\end{align}

\begin{figure}[tb]
  \centerline{
  \includegraphics[width=6cm]{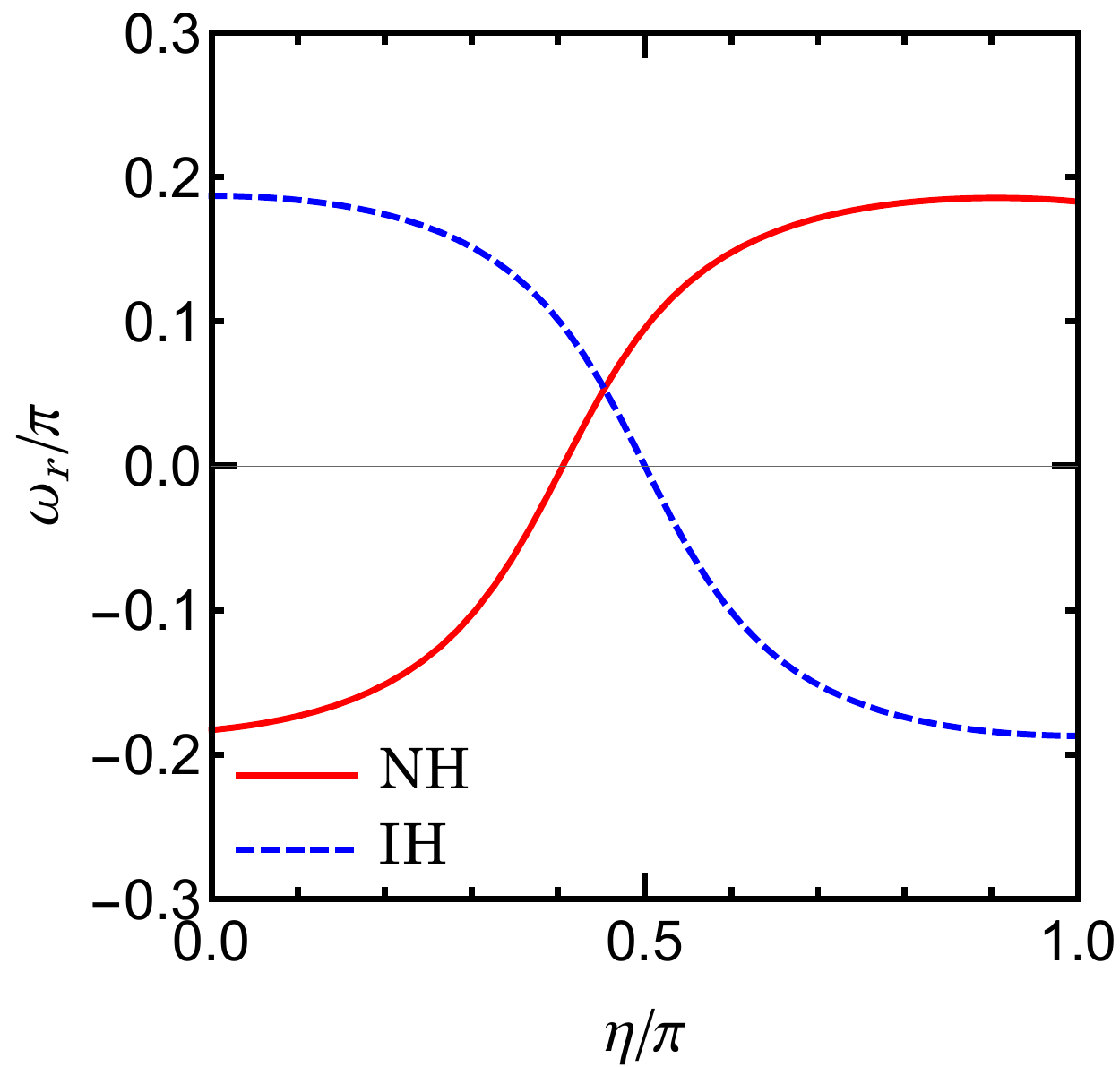}%
  \hspace{5mm}
  \includegraphics[width=5.85cm]{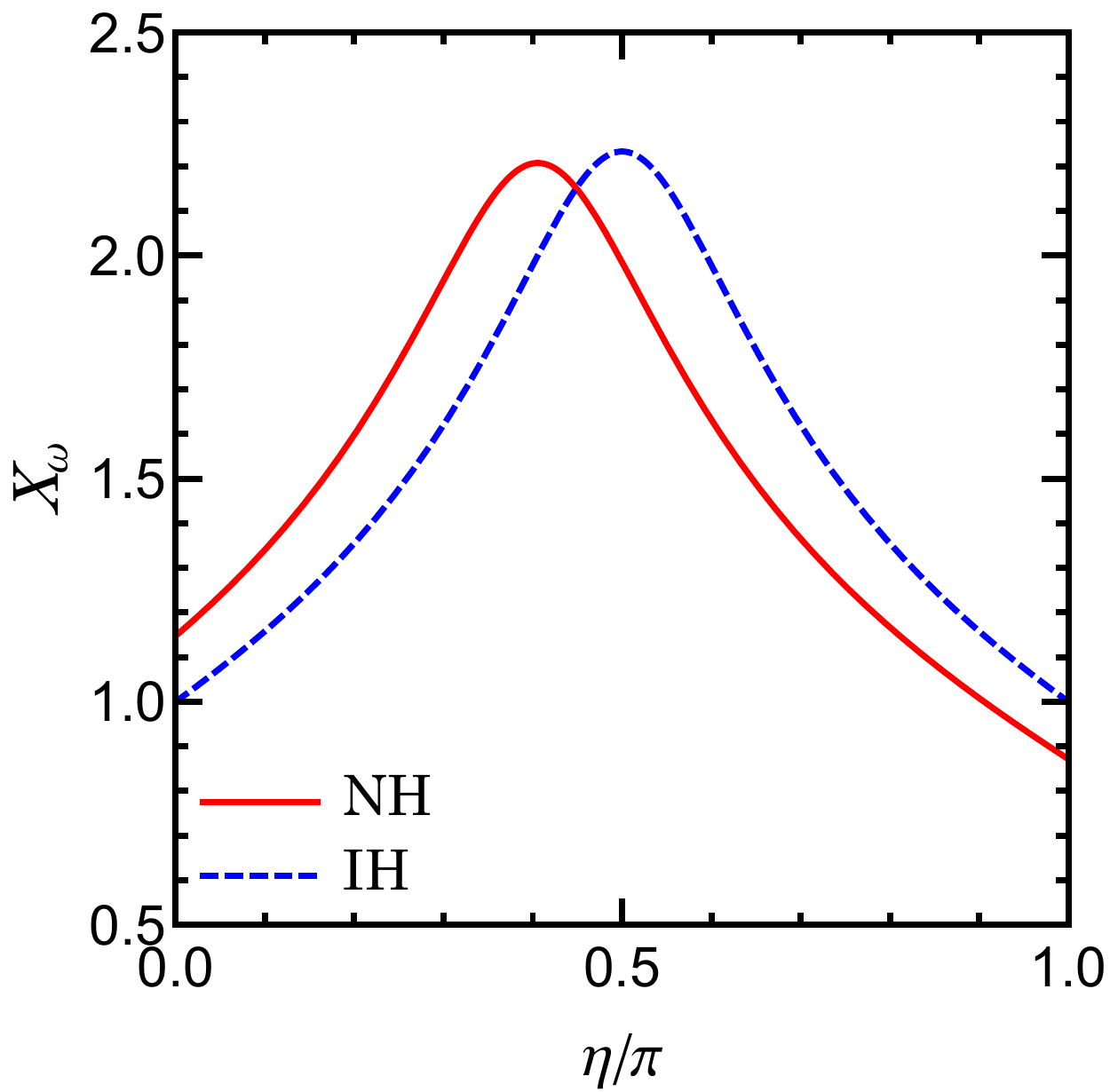}%
  }%
  \vspace{-2ex}
  \caption{
  	Required values of $\omega_r$ (left) and
		$X_\omega$ (right) for the vanishing effective mass
		in the NH or IH case (red solid or blue dashed line).
  }
  \label{fig:FIG_NH_OM}
\end{figure}
In Fig.~\ref{fig:FIG_NH_OM} we show the real and imaginary parts of $\omega$ 
satisfying the cancellation condition~(\ref{Eq:vanishing-cond}).
It is found from Eq.~(\ref{Eq:vanishing-cond}) that the maximal value of $X_\omega$ 
is achieved by the $CP$ violating phases
$\delta + \eta = -\pi/2$ for the NH case 
while $\eta=\pi/2$ in the IH case. 
Notice that $X_\omega$ becomes unity ({\it i.e.}, no imaginary part of $\omega$)
when $\eta=-\delta$ in the NH case and 
$\eta=0~(\pi)$ in the IH case, respectively.

\section{Discussions and conclusions}
\begin{figure}[t]
  \centerline{
  \hspace{-5mm}
  \includegraphics[width=6cm]{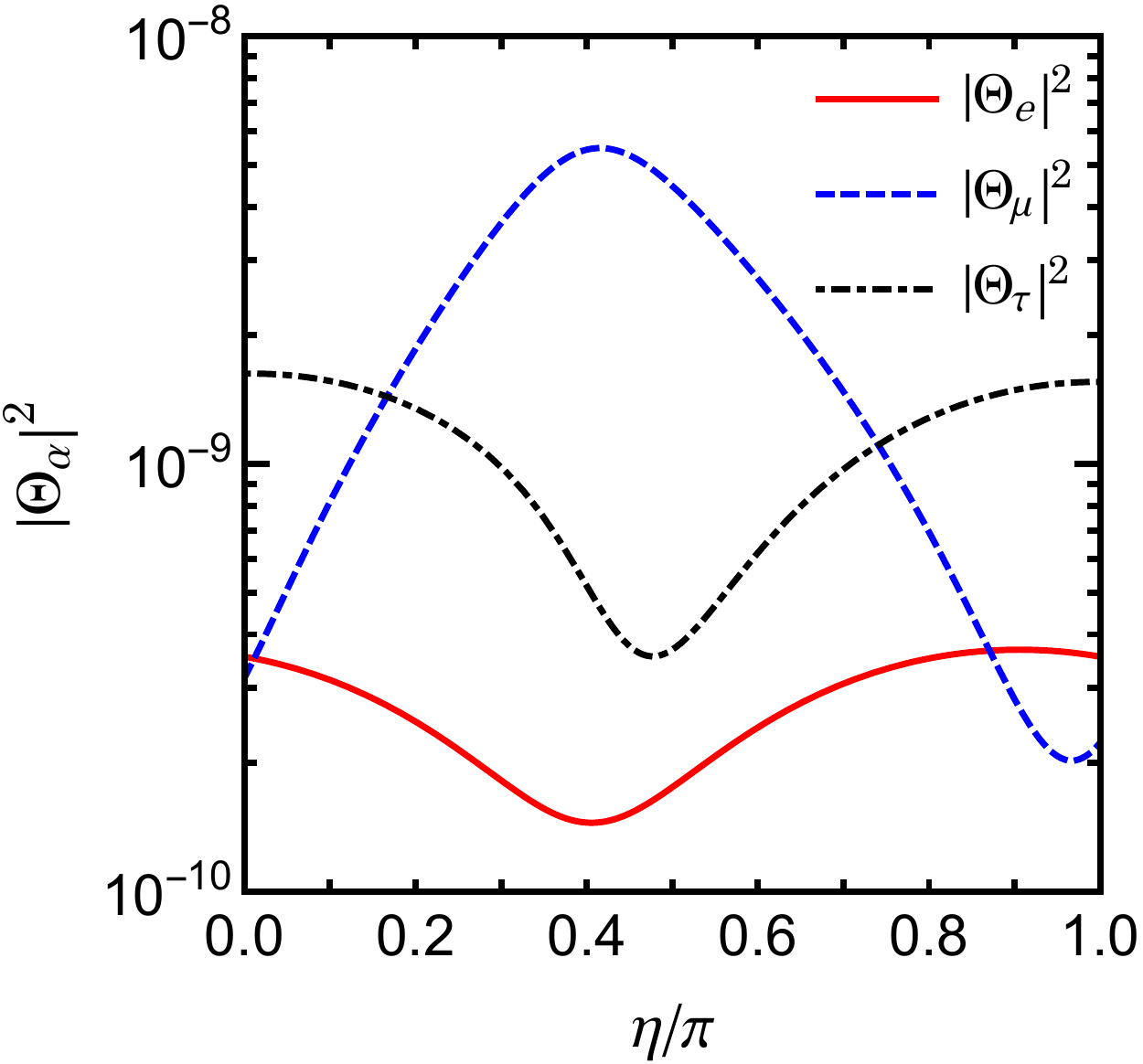}%
  \hspace{5mm}
  \includegraphics[width=6cm]{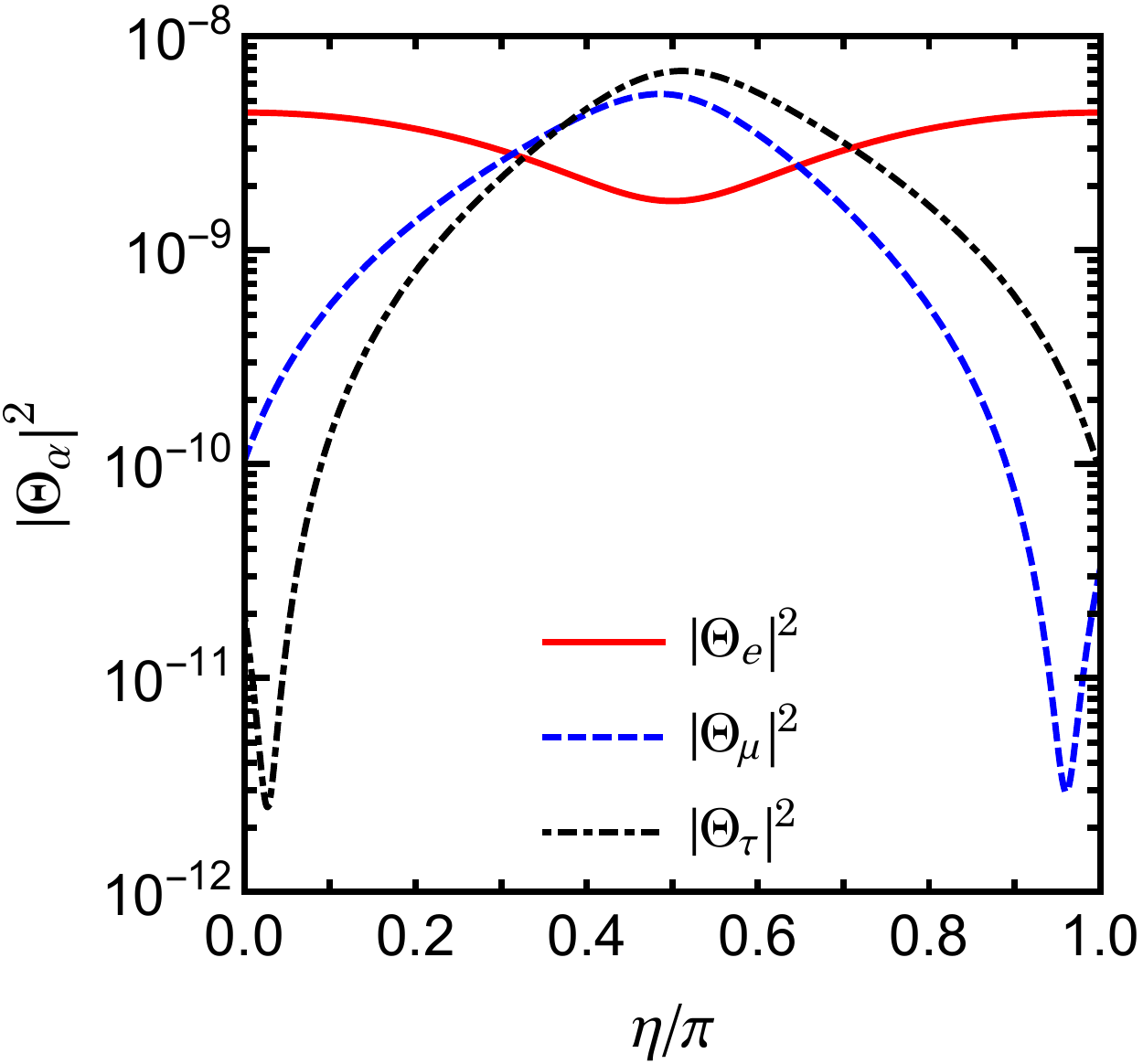}%
   }%
  \vspace{-2ex}
  \caption{
  	Mixing elements $|\Theta_{\alpha 1}|^2$ 
		for the vanishing effective neutrino mass
		in the NH (left) and IH (right) cases. 
		Here, $M_1 = 10$~MeV for both cases. 
		The solid (red), dashed (blue), and dot-dashed (black) lines represent 
		$\alpha = e$, $\mu$, and $\tau$, respectively. 
  }
  \label{fig:FIG_NH_THsq}
\end{figure}

It is, therefore, found that the contribution to $m_{\rm eff}$
from active neutrinos can be obscured by the light right-handed neutrino
when Eq.~(\ref{Eq:vanishing-cond}) is fulfilled. 
In this case, we can determine the mixing elements of of $N_1$
for a given mass depending on the Majorana phase. 
This point is illustrated in Fig.~\ref{fig:FIG_NH_THsq}. 

Interestingly, the flavor structure of the mixing elements
highly depends on the values of Majorana phase $\eta$. 
This pattern of the mixing elements may be tested by future 
direct search experiments of right-handed neutrinos. 
In the NH case
$|\Theta_{\mu 1}|^2 \gg |\Theta_{e 1}|^2$ 
and then the experiments like the peak search in $K \to \mu + N_1$, for instance,
would help the observation. 
On the other hand, in the IH case
$|\Theta_{e 1}|^2$ or $|\Theta_{\mu 1}|^2$ becomes dominant 
depending on $\eta$, and $K \to e + N_1$ or $K \to \mu + N_1$ would be
the golden channel for the discovery.
Since the relative sizes of the mixing elements are not so much identical
we can extract important information of the mass hierarchy 
and the Majorana phase $\eta$ 
from the combination of $|\Theta_{e 1}|^2$ and $|\Theta_{\mu 1}|^2$ 
under the situation of no $0 \nu \beta \beta$ decay is observed.

When the effective mass vanishes, the lifetime of $N_1$ can be
predicted by $M_1$ and $\eta$.  
In the mass region of interest the possible decay channels are
$N_1 \to \nu \nu \nu $ and $N_1 \to \nu e^+ e^-$ (when $M_1 > 2 m_e$) 
and we find the range of the lifetime is
\begin{align}
	\tau \simeq 
	\left\{
		\begin{array}{l}
		(5\mathchar`-6) \times 10^{7}~\mbox{sec} 
		\left( \displaystyle \frac{10~\mbox{MeV}}{M_1}\right)^4
		\hspace{5mm} \mbox{for the NH case}
		 \\[4ex]
		(0.9\mathchar`-2) \times 10^{7}~\mbox{sec}
		\left( \displaystyle \frac{10~\mbox{MeV}}{M_1}\right)^4 
		\hspace{2mm} \mbox{for the IH case}
		\end{array}
	\right. \,.
\end{align}
The suggested values of the lifetime are so long that 
$N_1$ decays after the onset of the big bang nucleosynthesis (BBN) 
and would destroy the success of the BBN and/or conflict with the observational
data of the cosmic microwave background radiation. 
One possibility to avoid this difficulty is the dilution of the $N_1$ abundance
by the late time entropy production.   Such an additional production may be
realized by the decay of the heavier right-handed neutrino $N_2$~\cite{Asaka:2006ek}. 

To summarize we have examined the neutrinoless double beta decay in 
the seesaw mechanism by two right-handed neutrinos.
The Majorana nature of active neutrinos and right-handed neutrinos breaks the lepton number of the theory, 
which may lead to the neutrinoless double beta decay. 
When the masses of right-handed neutrinos, however, are lighter than or comparable 
to the scale $\Lambda_\beta$, they can give a significant effect, 
and the effective mass can vanish in some cases.

In this paper we have found a possibility when
$M_1 \lesssim \Lambda_\beta \ll M_2$.
It has been shown that $N_1$ contribution can obliterate 
the neutrinoless double beta decay even if active neutrinos do contribute it.
If this is the case, the unknown parameters related to right-handed neutrinos appeared in $\Omega$ are highly restricted 
and the mixing elements of $N_1$ can be determined by the Majorana phase 
and the mass hierarchy of active neutrinos. 
Inversely speaking, we may obtain important information 
of the missing piece of neutrino properties, 
namely the Majorana phase and the mass hierarchy, 
from the relative sizes among the mixing elements of right-handed neutrinos 
measured at the future terrestrial experiments 
together with no neutrinoless double beta decay.

\section*{Acknowledgments}
The work of T.A. was partially supported by JSPS KAKENHI 
Grants No. 17K05410, No. 18H03708, No. 19H05097, and No. 20H01898.
The work of H.I. was supported by JSPS KAKENHI Grant No. 18H03708.



\begin{thebibliography}{100}
%
\bibitem{Pas:2015eia}
H.~Päs and W.~Rodejohann,
New J. Phys. \textbf{17} (2015) no.11, 115010
doi:10.1088/1367-2630/17/11/115010
[arXiv:1507.00170 [hep-ph]].

%
\bibitem{Arnaboldi:2002te}
  C.~Arnaboldi {\it et al.},
  Phys.\ Lett.\ B {\bf 557} (2003) 167
  doi:10.1016/S0370-2693(03)00212-0
  [hep-ex/0211071].


\bibitem{Umehara:2008ru}
  S.~Umehara {\it et al.},
  Phys.\ Rev.\ C {\bf 78} (2008) 058501
  doi:10.1103/PhysRevC.78.058501
  [arXiv:0810.4746 [nucl-ex]].


\bibitem{Barabash:2010bd}
  A.~S.~Barabash {\it et al.} [NEMO Collaboration],
  Phys.\ Atom.\ Nucl.\  {\bf 74} (2011) 312
  doi:10.1134/S1063778811020062
  [arXiv:1002.2862 [nucl-ex]].


\bibitem{Gando:2012zm}
  A.~Gando {\it et al.} [KamLAND-Zen Collaboration],
  Phys.\ Rev.\ Lett.\  {\bf 110} (2013) no.6,  062502
  doi:10.1103/PhysRevLett.110.062502
  [arXiv:1211.3863 [hep-ex]].


\bibitem{Agostini:2013mzu}
  M.~Agostini {\it et al.} [GERDA Collaboration],
  Phys.\ Rev.\ Lett.\  {\bf 111} (2013) no.12,  122503
  doi:10.1103/PhysRevLett.111.122503
  [arXiv:1307.4720 [nucl-ex]].


\bibitem{Albert:2014awa}
  J.~B.~Albert {\it et al.} [EXO-200 Collaboration],
  Nature {\bf 510} (2014) 229
  doi:10.1038/nature13432
  [arXiv:1402.6956 [nucl-ex]].


\bibitem{Andringa:2015tza}
  S.~Andringa {\it et al.} [SNO+ Collaboration],
  Adv.\ High Energy Phys.\  {\bf 2016} (2016) 6194250
  doi:10.1155/2016/6194250
  [arXiv:1508.05759 [physics.ins-det]].


\bibitem{Arnold:2015wpy}
  R.~Arnold {\it et al.} [NEMO-3 Collaboration],
  Phys.\ Rev.\ D {\bf 92} (2015) no.7,  072011
  doi:10.1103/PhysRevD.92.072011
  [arXiv:1506.05825 [hep-ex]].


\bibitem{Arnold:2016ezh}
  R.~Arnold {\it et al.} [NEMO-3 Collaboration],
  Phys.\ Rev.\ D {\bf 93} (2016) no.11,  112008
  doi:10.1103/PhysRevD.93.112008
  [arXiv:1604.01710 [hep-ex]].


\bibitem{KamLAND-Zen:2016pfg}
  A.~Gando {\it et al.} [KamLAND-Zen Collaboration],
  Phys.\ Rev.\ Lett.\  {\bf 117} (2016) no.8,  082503
   Addendum: [Phys.\ Rev.\ Lett.\  {\bf 117} (2016) no.10,  109903]
  doi:10.1103/PhysRevLett.117.109903, 10.1103/PhysRevLett.117.082503
  [arXiv:1605.02889 [hep-ex]].


\bibitem{Elliott:2016ble}
  S.~R.~Elliott {\it et al.},
  J.\ Phys.\ Conf.\ Ser.\  {\bf 888} (2017) no.1,  012035
  doi:10.1088/1742-6596/888/1/012035
  [arXiv:1610.01210 [nucl-ex]].


\bibitem{Arnold:2016qyg}
  R.~Arnold {\it et al.} [NEMO-3 Collaboration],
  Phys.\ Rev.\ D {\bf 94} (2016) no.7,  072003
  doi:10.1103/PhysRevD.94.072003
  [arXiv:1606.08494 [hep-ex]].


\bibitem{Arnold:2016bed}
  R.~Arnold {\it et al.} [NEMO-3 Collaboration],
  Phys.\ Rev.\ D {\bf 95} (2017) no.1,  012007
  doi:10.1103/PhysRevD.95.012007
  [arXiv:1610.03226 [hep-ex]].


\bibitem{Agostini:2017iyd}
  M.~Agostini {\it et al.},
  Nature {\bf 544} (2017) 47
  doi:10.1038/nature21717
  [arXiv:1703.00570 [nucl-ex]].


\bibitem{Aalseth:2017btx}
  C.~E.~Aalseth {\it et al.} [Majorana Collaboration],
  Phys.\ Rev.\ Lett.\  {\bf 120} (2018) no.13,  132502
  doi:10.1103/PhysRevLett.120.132502
  [arXiv:1710.11608 [nucl-ex]].


\bibitem{Albert:2017owj}
  J.~B.~Albert {\it et al.} [EXO Collaboration],
  Phys.\ Rev.\ Lett.\  {\bf 120} (2018) no.7,  072701
  doi:10.1103/PhysRevLett.120.072701
  [arXiv:1707.08707 [hep-ex]].


\bibitem{Alduino:2017ehq}
  C.~Alduino {\it et al.} [CUORE Collaboration],
  Phys.\ Rev.\ Lett.\  {\bf 120} (2018) no.13,  132501
  doi:10.1103/PhysRevLett.120.132501
  [arXiv:1710.07988 [nucl-ex]].


\bibitem{Agostini:2018tnm}
  M.~Agostini {\it et al.} [GERDA Collaboration],
  Phys.\ Rev.\ Lett.\  {\bf 120} (2018) no.13,  132503
  doi:10.1103/PhysRevLett.120.132503
  [arXiv:1803.11100 [nucl-ex]].


\bibitem{Azzolini:2018dyb}
  O.~Azzolini {\it et al.} [CUPID-0 Collaboration],
  Phys.\ Rev.\ Lett.\  {\bf 120} (2018) no.23,  232502
  doi:10.1103/PhysRevLett.120.232502
  [arXiv:1802.07791 [nucl-ex]].


\bibitem{Arnold:2018tmo}
  R.~Arnold {\it et al.},
  Eur.\ Phys.\ J.\ C {\bf 78} (2018) no.10,  821
  doi:10.1140/epjc/s10052-018-6295-x
  [arXiv:1806.05553 [hep-ex]].


\bibitem{Adams:2019jhp}
  D.~Q.~Adams {\it et al.} [CUORE Collaboration],
  Phys.\ Rev.\ Lett.\  {\bf 124} (2020) no.12,  122501
  doi:10.1103/PhysRevLett.124.122501
  [arXiv:1912.10966 [nucl-ex]].


\bibitem{Alvis:2019sil}
  S.~I.~Alvis {\it et al.} [Majorana Collaboration],
  Phys.\ Rev.\ C {\bf 100} (2019) no.2,  025501
  doi:10.1103/PhysRevC.100.025501
  [arXiv:1902.02299 [nucl-ex]].


\bibitem{Agostini:2019hzm}
  M.~Agostini {\it et al.} [GERDA Collaboration],
  Science {\bf 365} (2019) 1445
  doi:10.1126/science.aav8613
  [arXiv:1909.02726 [hep-ex]].


\bibitem{Azzolini:2019tta}
  O.~Azzolini {\it et al.} [CUPID Collaboration],
  Phys.\ Rev.\ Lett.\  {\bf 123} (2019) no.3,  032501
  doi:10.1103/PhysRevLett.123.032501
  [arXiv:1906.05001 [nucl-ex]].


\bibitem{Alenkov:2019jis}
  V.~Alenkov {\it et al.},
  Eur.\ Phys.\ J.\ C {\bf 79} (2019) no.9,  791
  doi:10.1140/epjc/s10052-019-7279-1
  [arXiv:1903.09483 [hep-ex]].

\bibitem{Anton:2019wmi}
  G.~Anton {\it et al.} [EXO-200 Collaboration],
  Phys.\ Rev.\ Lett.\  {\bf 123} (2019) no.16,  161802
  doi:10.1103/PhysRevLett.123.161802
  [arXiv:1906.02723 [hep-ex]].

\bibitem{Minkowski:1977sc}
  P.~Minkowski,
  Phys.\ Lett.\  {\bf 67B} (1977) 421.
  doi:10.1016/0370-2693(77)90435-X


\bibitem{Yanagida:1979as}
  T.~Yanagida,
  Conf.\ Proc.\ C {\bf 7902131} (1979) 95.


\bibitem{Yanagida:1980xy}
  T.~Yanagida,
  Prog.\ Theor.\ Phys.\  {\bf 64} (1980) 1103.
  doi:10.1143/PTP.64.1103
%
\bibitem{Ramond:1979}
P.~Ramond, 
in {\em Talk given at the Sanibel Symposium}, 
Palm Coast, Fla., Feb.~25-Mar.~2, 1979, preprint CALT-68-709
(retroprinted as hep-ph/9809459).

\bibitem{GellMann:1980vs} 
  M. Gell-Mann, P. Ramond, and R. Slansky, in Supergravity, edited by.P. van Niewwenhuizen and D. Freedman (North Holland, Amsterdam, 1979)
  [arXiv:1306.4669 [hep-th]].
  
\bibitem{Glashow:1979}
S.~L.~Glashow,
in {\em Proc. of the Carg\'ese  Summer Institute on Quarks and Leptons},
Carg\'ese, July 9-29, 1979, 
eds. M.~L\'evy et. al, , (Plenum, 1980, New York), p707.

\bibitem{Mohapatra:1979ia}
  R.~N.~Mohapatra and G.~Senjanovic,
  Phys.\ Rev.\ Lett.\  {\bf 44} (1980) 912.

%
\bibitem{Blennow:2010th}
M.~Blennow, E.~Fernandez-Martinez, J.~Lopez-Pavon and J.~Menendez,
JHEP \textbf{07} (2010), 096
doi:10.1007/JHEP07(2010)096
[arXiv:1005.3240 [hep-ph]].

\bibitem{Asaka:2011pb}
  T.~Asaka, S.~Eijima and H.~Ishida,
  JHEP {\bf 1104} (2011) 011
  doi:10.1007/JHEP04(2011)011
  [arXiv:1101.1382 [hep-ph]].


\bibitem{Akhmedov:1998qx}
E.~K.~Akhmedov, V.~A.~Rubakov and A.~Y.~Smirnov,
Phys. Rev. Lett. \textbf{81} (1998), 1359-1362
doi:10.1103/PhysRevLett.81.1359
[arXiv:hep-ph/9803255 [hep-ph]].

\bibitem{Asaka:2005pn}
T.~Asaka and M.~Shaposhnikov,
Phys. Lett. B \textbf{620} (2005), 17-26
doi:10.1016/j.physletb.2005.06.020
[arXiv:hep-ph/0505013 [hep-ph]].


\bibitem{Maki:1962mu}
  Z.~Maki, M.~Nakagawa and S.~Sakata,
  Prog.\ Theor.\ Phys.\  {\bf 28} (1962) 870.

\bibitem{Pontecorvo:1958}
  B.~Pontecorvo, Sov.\ Phys.\ JETP\ {\bf 7} (1958) 172.

\bibitem{Casas:2001sr}
  J.~A.~Casas and A.~Ibarra,
  Nucl.\ Phys.\ B {\bf 618} (2001) 171
  doi:10.1016/S0550-3213(01)00475-8
  [hep-ph/0103065].

\bibitem{Abada:2006ea}
  A.~Abada, S.~Davidson, A.~Ibarra, F.-X.~Josse-Michaux, M.~Losada and A.~Riotto,
  JHEP {\bf 0609} (2006) 010
  doi:10.1088/1126-6708/2006/09/010
  [hep-ph/0605281].

%
\bibitem{Esteban:2018azc}
I.~Esteban, M.~Gonzalez-Garcia, A.~Hernandez-Cabezudo, M.~Maltoni and T.~Schwetz,
JHEP \textbf{01} (2019), 106
doi:10.1007/JHEP01(2019)106
[arXiv:1811.05487 [hep-ph]].

%
\bibitem{nufit}
I.~Esteban, M.~Gonzalez-Garcia, A.~Hernandez-Cabezudo, M.~Maltoni and T.~Schwetz,
``NuFiT 4.1: Three-neutrino fit based on data available in July 2019,''
www.nu-fit.org.


\bibitem{Faessler:2014kka}
A.~Faessler, M.~González, S.~Kovalenko and F.~Šimkovic,
Phys. Rev. D \textbf{90} (2014) no.9, 096010
doi:10.1103/PhysRevD.90.096010
[arXiv:1408.6077 [hep-ph]].

%
\bibitem{Hyvarinen:2015bda}
J.~Hyvärinen and J.~Suhonen,
Phys. Rev. C \textbf{91} (2015) no.2, 024613
doi:10.1103/PhysRevC.91.024613
%
\bibitem{Barea:2015kwa}
J.~Barea, J.~Kotila and F.~Iachello,
Phys. Rev. C \textbf{91} (2015) no.3, 034304
doi:10.1103/PhysRevC.91.034304
[arXiv:1506.08530 [nucl-th]].

%
\bibitem{Menendez:2017fdf}
J.~Menéndez,
J. Phys. G \textbf{45} (2018) no.1, 014003
doi:10.1088/1361-6471/aa9bd4
[arXiv:1804.02105 [nucl-th]].


%
\bibitem{Asaka:2006ek}
  T.~Asaka, M.~Shaposhnikov and A.~Kusenko,
  Phys.\ Lett.\ B {\bf 638} (2006) 401
  doi:10.1016/j.physletb.2006.05.067
  [hep-ph/0602150].

\end{thebibliography}
\end{document}